# Adversarial VR: An Open-Source Testbed for Evaluating Adversarial Robustness of VR Cybersickness Detection and Mitigation

Istiak Ahmed *  
University of Missouri-Columbia

Ripan Kumar Kundu †  
University of Missouri-Columbia

Khaza Anuarul Hoque ‡  
University of Missouri-Columbia

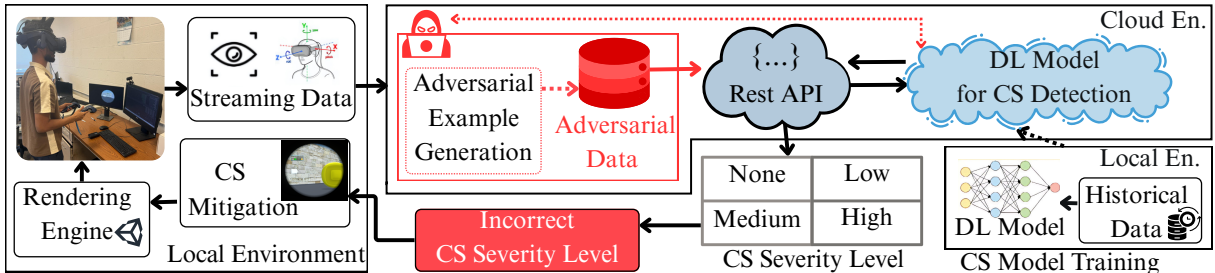

Figure 1: An overview of our developed *Adversarial-VR* testbed, consisting of two essential components: a local VR environment (En.) and a cloud-based VR environment. (Left) The local VR environment runs the maze simulation and performs on-device cybersickness (CS) mitigation. The cloud-based environment handles cybersickness detection using a deployed DL model, while the attacker generates adversarial examples. These adversarial examples are then fed into the DL models, altering the severity of cybersickness (e.g., none, low, medium, and high) and triggering incorrect mitigation strategies.

## ABSTRACT

Deep learning (DL)-based automated cybersickness detection methods, along with adaptive mitigation techniques, can enhance user comfort and interaction. However, recent studies show that these DL-based systems are susceptible to adversarial attacks; small perturbations to sensor inputs can degrade model performance, trigger incorrect mitigation, and disrupt the user's immersive experience (UIX). Additionally, there is a lack of dedicated open-source testbeds that evaluate the robustness of these systems under adversarial conditions, limiting the ability to assess their real-world effectiveness. To address this gap, this paper introduces *Adversarial-VR*, a novel real-time VR testbed for evaluating DL-based cybersickness detection and mitigation strategies under adversarial conditions. Developed in Unity, the testbed integrates two state-of-the-art (SOTA) DL models: DeepTCN and Transformer, which are trained on the open-source MazeSick dataset, for real-time cybersickness severity detection and applies a dynamic visual tunneling mechanism that adjusts the field-of-view based on model outputs. To assess robustness, we incorporate three SOTA adversarial attacks: MI-FGSM, PGD, and C&W, which successfully prevent cybersickness mitigation by fooling DL-based cybersickness models' outcomes. We implement these attacks using a testbed with a custom-built VR Maze simulation and an HTC Vive Pro Eye headset, and we open-source our implementation for widespread adoption by VR developers and researchers. Results show that these adversarial attacks are capable of successfully fooling the system. For instance, the C&W attack results in a $5.94\times$ decrease in accuracy for the Transformer-based cybersickness model compared to the accuracy without the attack.

---

*e-mail: ia5qq@missouri.edu  
†e-mail: rkundu@missouri.edu  
‡e-mail: hoquek@missouri.edu

## 1 INTRODUCTION

Virtual Reality (VR) promises unparalleled immersive experiences across various domains, from gaming [28] and education [37] to healthcare [42]. However, as VR technology advances, it introduces significant security vulnerabilities. Researchers have highlighted critical risks, such as security and privacy attacks (SPS) [37], network and GPU-based attacks [39], etc., which can compromise the integrity of VR systems and lead to disruptions in user experience by inducing cybersickness. User experience is a vital aspect of VR, and cybersickness remains a significant obstacle to the broader acceptance of VR. Several machine learning (ML) and deep learning (DL) methods have been proposed for the automatic detection and mitigation of cybersickness to improve user comfort and safety [16, 17, 20]. In an ideal setup (without adversarial interference), these ML/DL models automatically detect the severity of cybersickness from VR sensor data and trigger adaptive mitigation techniques, such as narrowing the dynamic field of view (FOV) [7, 15, 20]. However, ML/DL algorithms are vulnerable to carefully crafted adversarial examples [24], which also applicable for the VR cybersickness use cases [18]. This shows the importance of analyzing and evaluating the robustness of DL models when used for automatic cybersickness detection and mitigation.

Despite the growing importance of ML/DL-driven cybersickness detection and mitigation, there is a significant gap in the availability of open-source testbeds capable of evaluating these real-time ML/DL-based detection methods. Furthermore, the research community has yet to develop comprehensive testbeds that automatically implement and assess the entire pipeline for cybersickness detection and mitigation using ML/DL techniques. While some testbeds exist [4, 11, 35, 43], they do not support the automatic integration of detection and mitigation systems. Additionally, there is a lack of testbeds designed to evaluate the adversarial robustness of these AI models in the context of cybersickness detection and mitigation. This gap in available resources and evaluation frameworks underscores the need for further research and the development of an open-source testbed that not only supports real-time ML/DL-based detection but also evaluates the resilience of these systems against adversarial attacks, thereby motivating the importance of this work.

Motivated by the above-mentioned limitation of existing works, this paper introduces *Adversarial-VR*, a novel real-time VR testbed for evaluating DL-based automatic cybersickness detection and mitigation strategies under adversarial attack conditions, as shown in Figure 1. To the best of our knowledge, it is the first open-source testbed to assess the robustness of DL-based automatic cybersickness detection and mitigation systems against adversarial attack conditions. Our key contributions are as follows:

- We develop our testbed by incorporating two state-of-the-art (SOTA) DL models: Deep Temporal Convolutional Network (DeepTCN) and the Transformer. It is worth mentioning that any DL model can be integrated into our proposed testbed. These models are trained using MazeSick, a publicly available SOTA VR cybersickness dataset [20]. Upon detecting cybersickness, the system automatically triggers mitigation techniques using Unity's *Tunneling Vignette* system [36]. Specifically, we implement a dynamic field-of-view (FOV) adjustment technique that adapts to the severity of the user's symptoms. Our evaluation confirms that the system performs effectively under ideal conditions, i.e., when no adversarial attacks are present, as shown in Figure 2.

- Our testbed supports generating adversarial examples and injecting them into the cybersickness detection models, thereby manipulating the outcome of cybersickness detection. To craft these adversarial inputs, we employ three widely used attack algorithms: Momentum Iterative Fast Gradient Sign Method (MI-FGSM) [8], Projected Gradient Descent (PGD) [23], and Carlini-Wagner (C&W) [5] method. It is important to note that any adversarial example generation algorithms can also be integrated into our testbed. Our testbed evaluation covers both white-box and black-box attack scenarios. Results using the MazeSick dataset demonstrate that the proposed adversarial approach can effectively deceive the detection system. For instance, the C&W attack results in a 5.94× drop in detection accuracy for the Transformer-based cybersickness detection model compared to its accuracy without adversarial attacks. This manipulation prevents the activation of cybersickness mitigation mechanisms, significantly degrading the UIX.

- Finally, to support widespread adoption by VR developers and researchers, we release our system as an open-source testbed. By making our implementation publicly available [1], we aim to foster community-driven experimentation and advancement in adversarial robustness for VR. The testbed is implemented within a custom-built VR Maze simulation, utilizing the HTC Vive Pro Eye headset, which provides a realistic and extensible environment for evaluation.

## 2 RELATED WORKS

VR systems have emerged as critical platforms for immersive experiences across various domains, yet they remain vulnerable to sophisticated security threats that compromise both system integrity and user safety. Recent research has identified diverse attack vectors targeting VR environments, including security and privacy attacks (SPS) [37], network and GPU based attacks [39], and sophisticated manipulation techniques that exploit the sensory-rich nature of virtual environments. These vulnerabilities become particularly concerning when considering cybersickness, a fundamental challenge that significantly impacts VR adoption and user experience [34, 40]. Cybersickness manifests through symptoms including nausea, disorientation, and visual discomfort, creating substantial barriers to widespread VR acceptance. To address these challenges, researchers have proposed numerous ML and DL methods for automatic cybersickness detection and mitigation [16, 17, 19, 20, 22, 29, 33], leveraging multimodal sensor data from head and eye tracking to predict severity levels and trigger adaptive techniques such as dynamic field-of-view reduction [7, 9, 15, 34]. For instance, the authors in [20] utilized a DL-based transformer network on multimodal time-series VR data to detect cybersickness and implement adaptive mitigation techniques (e.g., dynamic FoV adjustments, Gaussian blurring) to alleviate its effects. However, this work does not address the potential vulnerabilities of ML/DL-based cybersickness detection methods to adversarial threats. Recent studies have shown that these ML/DL algorithms are vulnerable to adversarial attacks [24]. Adversarial attacks compromise the DL-based models prediction algorithm's ability to accurately detect and mitigate cybersickness, resulting in distorted user experiences and ultimately undermining trust in VR systems. Very recently, Kundu et al. [18] introduced a new type of VR attack, namely a cybersickness attack, by applying adversarial attacks. This attack successfully prevents the triggering of cybersickness mitigation by fooling DL-based cybersickness detection models and significantly hinders the UIX. However, a key limitation of their work is the absence of a realistic testbed, which is vital for VR developers and researchers. These testbeds play a crucial role in assessing the robustness of DL models for automated cybersickness detection and mitigation, ensuring the systems can effectively resist adversarial attacks and provide reliable user protection.

---

[1] https://github.com/dependable-cps/Adversarial-VR-Testbed

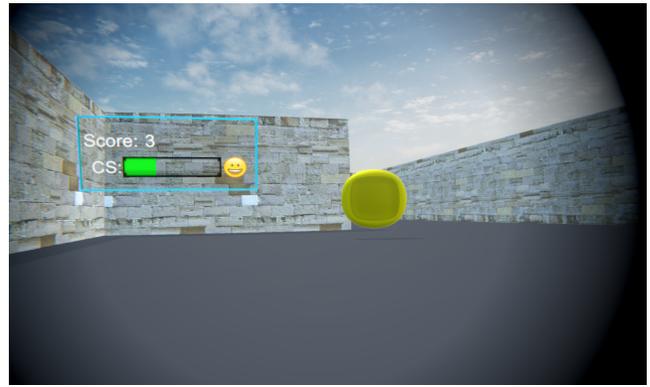

Figure 2: A first-person view within the testbed shows the VR user's coin collection score and the DL-enabled, real-time detection of cybersickness (CS) severity in the VR maze simulation. Upon detection, the tunnel effect dynamically adjusts the FOV according to the severity level, enabling real-time mitigation to enhance user comfort.

While existing research has made significant strides in detecting and mitigating cybersickness, current evaluation frameworks suffer from substantial limitations that hinder a comprehensive assessment of system robustness. Although several VR testbeds and frameworks have been proposed to study and alleviate cybersickness, each has limitations [4, 13, 31, 32]. For instance, BioVRSea et al. [31] combined a VR seasickness scenario with a moving platform and biomedical sensors to analyze motion-induced sickness, yet it targets a particular use-case (e.g, maritime motion) and does not incorporate real-time mitigation. While authors in [4] introduced the cybersickness evaluation testbed (CET-VR), which provides multiple immersive scenarios to elicit cybersickness and allows researchers to compare different mitigation techniques. However, their work does not provide an automatic cybersickness detection and mitigation technique by incorporating an ML/DL method. Indeed, Rouhani et al. [32] proposed a testbed to tackle VR sickness by incorporating different mitigation strategies (e.g., dynamic FOV reduction, blurring, or adding a virtual nose). Unfortunately, their testbed lacks an integrated, real-time ML detection mechanism to trigger the mitigation technique automatically. However, most of the prior works do not support the automatic ML/DL-based integration of detection and mitigation systems. Moreover, there is a noticeable absence of testbeds specifically designed to assess the adversarial robustness of AI models within the context of cybersickness detection and mitigation. This lack of resources and evaluation infrastructure underscores the need for continued research and the development of a comprehensive, open-source testbed that facilitates real-time ML/DL-based detection, while also evaluating system resilience against adversarial attacks, highlighting the significance of this work.

## 3 THREAT MODEL

The threat model we consider for adversarial attack in this paper is inspired by [18]. We consider two threat scenarios for adversarial attacks in DL-enabled automatic cybersickness detection and mitigation tasks: a white-box and a black-box attack scenario. In both scenarios, the parameters of the cybersickness detection model remain unchanged; instead, we rely solely on crafted adversarial examples to deceive the model. In the white-box scenario, we assume the adversary has complete knowledge of the cybersickness detection model and real-time access to the sensor data stream, enabling them to craft highly effective adversarial examples. In contrast, in the black-box scenario, the adversary lacks direct knowledge of the model's internal parameters or architecture. Instead, the attacker must make informed guesses about the cybersickness detection model (e.g., training a surrogate model) and rely on the transferability [41] of adversarial examples to fool the target system. In both cases, the attacker does not alter the mitigation engine or model parameters; instead, they modify the sensor inputs to induce misclassification.

**Attack Objective and Consequences:** The attacker's primary goal is to deceive the cybersickness detection model by causing it to misclassify the user's detected cybersickness level, thereby preventing the VR system from triggering the necessary mitigation or causing it to trigger inappropriate actions. This can result in two potential situations: (1) VR cybersickness mitigation fails to activate when needed, such as when the cybersickness detection model classifies the user's cybersickness level as *"none,"* preventing mitigation despite the user experiencing severe cybersickness, and (2) an incorrect type of VR cybersickness mitigation is triggered, mismatched with the severity level of cybersickness. The consequences of these cybersickness misclassifications on the user's experience are summarized in Table 1. For instance, if the actual cybersickness level is *"none"* but the model predicts *"low,"* or *"medium,"* or *"high"* due to the attack, the system will unnecessarily apply visual tunneling. This can diminish the user's immersion in the VR environment and potentially cause discomfort, even in the absence of cybersickness. Conversely, if the actual cybersickness level is *"high"* but the model predicts *"none,"* or *"low,"* or *"medium,"* the mitigation may be inadequate, leading to heightened discomfort and possibly prompting the user to abandon the VR session prematurely.

**Attack Surface:** In our testbed, we assume an adversary can gain access to the VR headset, the cloud storage containing the cybersickness detection model, or intercept communication between the headset and the cloud server, allowing them to execute malicious code. Such attacks may be carried out through methods like spoofing, phishing, evil twin attacks, or social engineering [30]. Additionally, the attack surface extends to the VR system's sensor data acquisition and processing pipeline, which includes real-time data streams from head and eye tracking sensors. Since the cybersickness detection model heavily relies on these multivariate time-series inputs, any manipulation of the sensor data can result in inaccurate classification outcomes. An adversary can exploit vulnerabilities in the software layer responsible for processing and transmitting this data, including the VR runtime, SDK middleware, or peripheral sensor APIs. Furthermore, side-channel vulnerabilities, such as timing patterns or GPU telemetry, can provide adversaries with indirect access to system behavior, allowing them to inject adversarial inputs stealthily. Even in closed-loop systems without internet connectivity, an adversary with local access to device firmware or runtime configurations can tamper with sensor calibration or directly manipulate data before it is processed. These manipulations can severely distort cybersickness detection, underscoring the critical need to secure the sensor-to-inference pipeline and prevent improper mitigations that can exacerbate user discomfort.

Table 1: Consequences of misclassification due to adversarial attacks

| Cybersickness | Predicted Level | Consequence |
|---|---|---|
| None | Low / Medium / High | Unnecessary tunneling |
| Low | None | Mitigation not triggered |
| Low | Medium / High | Over-mitigation |
| Medium | None / Low | Under-mitigation |
| Medium | High | Excessive mitigation |
| High | None / Low / Medium | Insufficient mitigation |

## 4 PROPOSED ADVERSARIAL-VR TESTBED DEVELOPMENT

In this section, we present the details of our proposed *Adversarial-VR* testbed, as illustrated in Fig. 1, which consists of three essential components: (1) cybersickness detection model (cloud environment), (2) cybersickness mitigation (local environment), and (3) adversarial attack module (cloud environment). Specifically, the cloud-based environment hosts the deployment of the DL-enabled cybersickness detection model and executes the adversarial attacks. In contrast, the local VR environment runs the VR simulation and includes on-device cybersickness mitigation. Our proposed testbed is implemented as a Unity project, designed to be compatible with VR headsets equipped with eye and head tracking capabilities (e.g., HTC VIVE Pro). The details of the Adversarial-VR testbed components are described below.

### 4.1 Cybersickness Detection Model (Cloud Environment)

For effective mitigation, it is crucial to detect the cybersickness accurately. Thus, we develop an accurate method for detecting cybersickness as a prerequisite for mitigation. To build an accurate cybersickness detection method, we utilize two SOTA DL models: Deep Temporal Convolutional Network (DeepTCN) and Transformer. We choose these DL models since they are popular and commonly used in the SOTA cybersickness detection research [16, 17, 20, 21]. The DeepTCN model consists of four layers that apply filters across a sliding window of input data. Each temporal convolutional network layer, featuring exponentially increasing dilations, is adept at capturing long-term dependencies in the data. This model is enhanced with residual and skip connections and predicts multiple tasks. In constructing our DeepTCN, we use five residual blocks, adopt Rectified Linear Unit (ReLU) as the activation function, and implement weighted batch normalization, where each layer is followed by a dropout layer with a dropout of 20% to prevent overfitting. A Transformer is a sequence-to-sequence architecture consisting of an encoder and a decoder [38]. The encoder takes the input sequence and maps it into a higher-dimensional vector, which is then fed into the decoder to generate an output sequence. In our work, the Transformer network consists of three sub-networks: (1) an embedding network, also known as a positional encoding (PE), that changes the dimension of input data and reflects positional information; (2) the encoder that trains the importance of data characteristics of time-series data through multi-head self-attention; and (3) the output layer that predicts cybersickness severity level through an activation function. We use a 64-dimensional attention vector, which goes through an attention process in the Transformer encoder. After that, we use an addition and normalization layer. Lastly, the MLP network is applied to capture the latent feature, which comprises 256 dimensions in the feed-forward layer. Both the DeepTCN and Transformer models operate on a fixed input window of 90 timesteps and output class probabilities for four severity categories via a softmax-activated dense layer. We use categorical cross-entropy in our DL-based cybersickness classification models as the loss function. In addition, we use $K$ fold cross-validation technique to train and validate the performance of the DL-based cybersickness detection models in which the dataset is partitioned into $k$ groups (i.e., in our case, $k = 10$). In this method, only one partition out of $k$ is used to test the model, while the remaining partitions are used to train the model. The method is repeated $k$ times, with each iteration selecting a new test partition and the remaining $(k-1)$ partitions as a training dataset to eliminate bias.

After training and validation, the developed DL-based cybersickness detection model is converted to Keras format and deployed in VR headsets, allowing for real-time inference. During inference, the model predicts cybersickness severity (e.g., none, low, medium, high) using real-time streaming VR simulation data (e.g., eye-tracking and head-tracking). These automatic cybersickness predictions are sent to the mitigation phase, which activates the appropriate FOV-based mitigation in the Unity environment, ensuring that correct mitigation triggers are effectively maintained to maintain user comfort during immersive experiences. It is important to note that no mitigation is triggered when the severity is classified as none.

### 4.2 Cybersickness Mitigation (Local Environment)

To address cybersickness in VR, we implement a dynamic visual tunneling technique that adaptively restricts the user's peripheral vision based on predicted discomfort severity, as supported by prior research demonstrating that dynamic field-of-view (FOV) modification can significantly reduce cybersickness while preserving user presence [9, 34]. Our system leverages Unity's Tunneling Vignette Controller, extended with a custom interface that dynamically maps cybersickness severity levels (e.g., none, low, medium, high) to predefined settings of Aperture Fraction (AF) and Feathering (F), as shown in Figure 4. AF controls the size of the Inner Field-of-View (IFOV), while F modulates the softness of the transition to the Outer Field-of-View (OFOV). As cybersickness severity escalates, AF and F are progressively reduced, narrowing the visible aperture and sharpening the peripheral cutoff, thereby minimizing peripheral optical flow to alleviate discomfort without significantly compromising spatial awareness, as detailed in Table 2 and visually illustrated in Figure 3. AF and F values for each severity level are determined based on the empirical guidelines presented by Fernandes and Feiner [9] and Teixeira and Palmisano [34], who outline effective FOV restriction ranges for reducing motion sickness while preserving the sense of presence. This approach enables a dynamic response to varying levels of user discomfort, ensuring clear visibility in the central visual field. Our implementation follows Unity's recommended comfort guidelines [36], utilizing shader materials for seamless, real-time, frame-by-frame adaptation based on the user's discomfort prediction. The system also includes fallback configurations for invalid inputs, enhancing robustness. The adaptive tunneling mechanism is visually demonstrated in Figure 3, which highlights the progressive restriction of peripheral vision across severity levels. The corresponding AF and F values, along with their visual effects, are comprehensively presented in Table 2. This integrated approach not only mitigates cybersickness but also enhances user immersion by maintaining a balance between comfort and spatial presence in VR environments.

### 4.3 Adversarial Attacks Module (Cloud Environment)

This section discusses the generation of adversarial examples and the execution of attacks within the developed testbed, aimed at deceiving the DL-enabled cybersickness detection model in real-time. Such attacks can potentially result in inappropriate mitigation triggers during VR simulations.

#### 4.3.1 Adversarial Example Generation

To evaluate the robustness of our DL-enabled cybersickness detection system, we apply three popular and effective SOTA adversarial attack algorithms: Momentum Iterative Fast Gradient Sign Method (MI-FGSM) [8], Projected Gradient Descent (PGD) [23], and Carlini-Wagner (C&W) [5]. We select these attacks since they are popular and commonly used in the SOTA DL-based cybersickness detection and time-series DL model robustness evaluations research [14, 18, 24]. It is important to note that we choose MI-FGSM instead of FGSM because it generates more stealthy and effective adversarial examples by stabilizing gradient updates through momentum [8]. These attacks induce incorrect detection of cybersickness severity, which in turn disrupts the activation of appropriate visual mitigation strategies, such as dynamic tunneling. It is worth noting that traditional adversarial attacks in computer vision, which manipulate pixel-level inputs, are not directly applicable to VR sensor data due to the temporal nature of time-series classification. In VR, the goal is to introduce subtle perturbations to integrated sensor data (e.g., eye and head tracking) to mislead DL-enabled cybersickness detection. To address this, we adapt vision-based adversarial algorithms (e.g., MI-FGSM, PGD, C&W) for time-series data, following techniques from [24]. Unlike static image attacks, our approach considers temporal dependencies and operates during immersive sessions to induce erroneous predictions and trigger unintended mitigation responses. These perturbations target features such as gaze direction, pupil metrics, and head orientation without modifying system internals. Note that sensors in VR headsets, such as eye-tracking and head-tracking sensors, capture measurements at regular intervals, forming multivariate time series (MTS) data.

Table 2: Mitigation parameters for the dynamic FOV adjustment based on cybersickness severity. The Aperture Fraction (AF) and Feathering (F) values adjust the IFOV and the OFOV transition, respectively, as cybersickness severity levels increase. Lower values of AF and F indicate stronger mitigation for more severe cybersickness (i.e., narrowing the FOV), while higher values represent a more gradual reduction in peripheral vision.

| Severity Level | Control Parameters | | IFOV-OFOV | Visual Effect |
|---|---|---|---|---|
| | AF | F | | |
| 0 (None) | 1.0f | 0.0f | Full (180°+) | No tunnel |
| 1 (Low) | 0.88f | 0.4f | 120°–155° | Wide tunnel |
| 2 (Medium) | 0.72f | 0.2f | 58°–110° | Moderate tunnel |
| 3 (High) | 0.52f | 0.05f | 36°–80° | Narrow tunnel |

#### 4.3.2 Adversarial Attack Execution

After gaining access to the VR device, the cloud-based cybersickness detection model, or the communication layer handling time-series VR sensor data transmission (e.g., eye and head tracking), an adversary executes the adversarial attack by injecting small, strategically designed perturbations into the real-time sensor stream, using the techniques described in Section 4.3.1. These streams, consisting of multivariate time series data from eye and head tracking, serve as input to the detection model. The attacker subtly alters the data stream in a way that remains imperceptible to users but disrupts DL-enabled cybersickness model predictions. These manipulated inputs are fed into the cybersickness detection model at runtime, potentially leading to the misclassification of sickness severity, where actual symptoms remain unmitigated, or false positives, where mitigation strategies such as visual tunneling are triggered unnecessarily. For instance, minor perturbations in head-tracking sequences can cause the model's output to shift from low to high severity, resulting in excessive FOV restriction and a degraded user immersive experience.

## 5 DATASET & EXPERIMENTAL SETUP

This section provides an overview of the dataset and the experimental setup to validate our proposed Adversarial-VR testbed.

### 5.1 Dataset

We train the DL models using SOTA open-source *MazeSick* dataset [20], a multimodal time-series dataset collected from 60

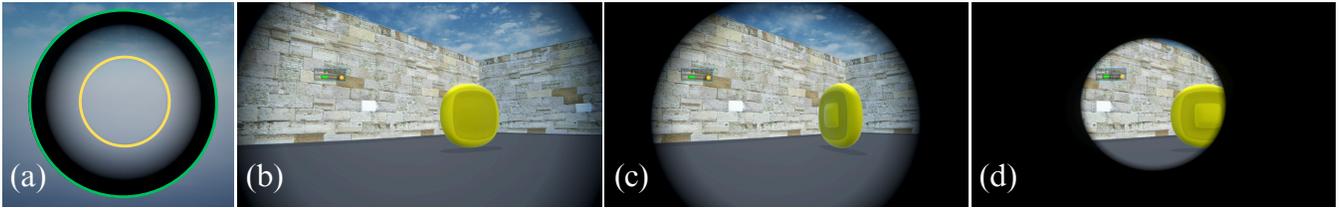

Figure 3: User's perspective within the developed testbed (Maze simulation) during cybersickness mitigation: **(a)** shows the adjustable IFOV (yellow circle) and OFOV (green circle) regions, which adapt based on the severity of the user's cybersickness. **(b)** Low severity - the FOV is moderately reduced, providing a slight tunnel effect. **(c)** Medium severity - the FOV is further restricted, increasing discomfort reduction while maintaining some peripheral vision. **(d)** High severity - the FOV is most restricted, offering significant reduction in peripheral vision to alleviate cybersickness, with sharper transitions in the outer FOV.

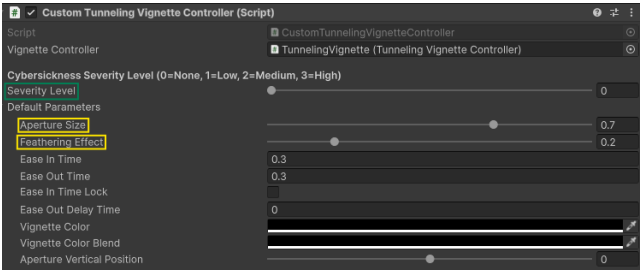

Figure 4: Custom tunneling vignette controller interface showing severity level mapping to aperture size and feathering parameters. The controller extends Unity's base implementation with automatic IFOV-OFOV configuration based on cybersickness severity.

participants (39 male, 21 female; age 18–38 years, mean = 23.6, SD = 3.21) representing Asian, Black, and Hispanic ethnicities. Data were captured during 12-minute VR maze navigation tasks using an HTC Vive Pro Eye headset, where participants employed smooth locomotion to collect coins in a dungeon-style environment. Eye-tracking features (e.g., left/right pupil diameter, normalized gaze direction/origin, etc.,) and head-tracking features (e.g., quaternion rotation) were recorded at 90 Hz via the Tobii HTC Vive Devkit [1]. The dataset includes Fast Motion Sickness Scale (FMS) scores (0-10) as ground truth. To support cybersickness classification, we conducted a distribution analysis of the FMS scores, following the approach in prior studies [16]. Specifically, FMS values were categorized into four severity levels: *none*, *low*, *medium*, and *high*.

### 5.2 Data Preprocessing

Initially, we performed outlier detection and removed all outliers that were more than three standard deviations away from the mean value using the z-score analysis method [6]. We then utilized exponential smoothing [12] to process the collected time series data, similar to [20], which effectively reduces the noise while minimizing deviations from the original signal. Finally, to ensure comparability and consistency between different measurement data (e.g., eye-tracking and head-tracking), we normalized [25] them using the following formula: $X_{normalized} = \frac{X-\mu}{\sigma}$. Here, $X_{normalized}$ represents the normalized sample, $\mu$ is the mean of the sample, and $\sigma$ is the standard deviation of the sample. Note that all the samples are time-synchronized, thus preserving the time dimension of the dataset.

### 5.3 Hyperparameters and Performance Metrics

We optimized our DL models using the Adam optimizer with a learning rate of 0.001, a batch size of 256, and 300 epochs. To prevent model overfitting, an early-stopping strategy with a patience of 30 epochs was employed during training. Furthermore, the performance of the DL-enabled cybersickness detection models (without adversarial attacks) is evaluated using standard metrics, such as accuracy, precision, recall, and F1-score, following established practices in prior work [22]. Similarly, the effectiveness of adversarial attacks is evaluated using the same set of metrics, consistent with the approach adopted in [10].

### 5.4 Experimental Setup

To validate our proposed Adversarial-VR testbed, we used the VR Maze simulation, similar to the one used to collect data for building the MazeSick dataset [16]. We use TensorFlow 2.4 [2] to train and evaluate the DeepTCN model, and PyTorch 2.3 [27] for the Transformer model. All models are trained on an Intel Core i9 Processor and 128GB RAM with an NVIDIA GeForce RTX 3090 Ti GPU. For implementing the dynamic FOV reduction mitigation technique, we utilized Unity's *Tunneling Vignette* system [36], which was extended with a custom controller interface that maps predicted severity levels to shader parameters: aperture fraction (AF) and feathering (F). Furthermore, to assess the robustness of our cybersickness detection models, we implemented MI-FGSM, C&W, and PGD adversarial attacks using the CleverHans library [26]. The HTC SRanipal SDK and Tobii HTC Vive Devkit API [1] were employed to capture precise VR simulation data (e.g., eye-tracking and head-tracking data). Finally, the VR simulation was displayed using an HTC Vive Pro Eye HMD. The server runs a Python Flask backend equipped with pre-trained DL models. It receives real-time eye and head movement data from the headset via a REST API and returns predictions of cybersickness severity (e.g., none/low/medium/high). It is important to note that we used the same CPU to manage the real-time data streaming, DL model inference, and VR simulation rendering.

### 5.5 Virtual Environment and Apparatus

The VR environment was a Unity-based maze navigation task built with OpenVR, featuring colorful brick walls, an open roof, and a cloudy sky backdrop, which was connected to a local edge server. The maze used a $15 \times 14$ grid layout covering 150m×140m (Unity scale: *Vector3(150, 1, 140)*), with each platform measuring 10m×10m. Designed for single-player use, it supported both room-scale and smooth locomotion. Users navigated the maze using a smooth locomotion technique under varying cognitive task conditions. The maze features turn-based segments and randomized paths, designed to induce cybersickness through unpredictable motion patterns. To enhance interactivity, we integrate a coin collection mechanic that encourages continuous navigation and natural head movement, increasing the likelihood of motion-induced discomfort. An HTC Vive Pro Eye VR headset with integrated eye-tracking and head-tracking was utilized, providing a resolution of 1440 × 1600 pixels per eye at a 90 Hz refresh rate and a 110-degree field of view.

## 6 EXPERIMENTAL RESULTS

This section presents the experimental results of our cybersickness detection and adversarial attack evaluation on the testbed.

**Cybersickness Detection (without Adversarial attacks):** Table 3 presents the performance of cybersickness severity classification in terms of accuracy, precision, recall, and F1-score using DeepTCN and Transformer models on the MazeSick dataset. We observe that the Transformer and DeepTCN models achieve classification accuracies of 95% and 91%, respectively. Overall, the Transformer model outperforms DeepTCN model in terms of precision, recall, and F1-score for cybersickness classification.

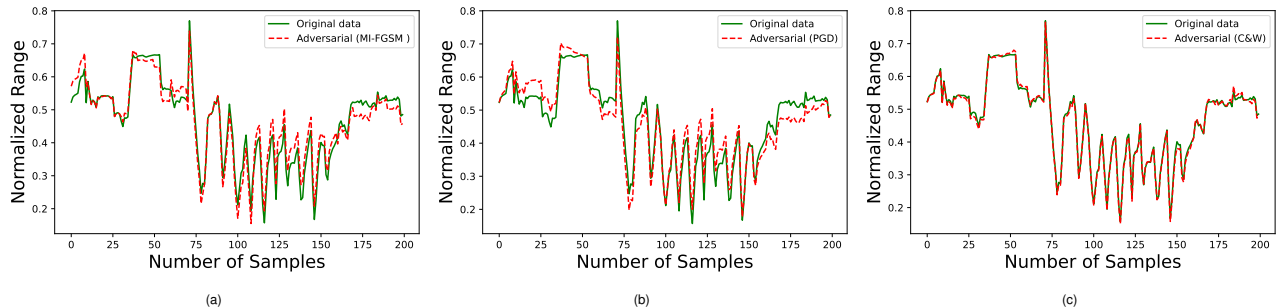

Figure 5: Visualization of original vs. adversarial VR sensor data streams under the Transformer model for different attack types. Solid green lines represent clean signals, while dashed red lines indicate adversarially perturbed inputs. Attacks target eye movement data as follows: **(a)** MI-FGSM, **(b)** PGD, and **(c)** C&W adversarial attacks.

Table 3: Mean accuracy (Acc.), precision, recall, and F1-score (F1-S) of cybersickness detection without adversarial attack models, and with MI-FGSM ($\varepsilon = 0.1$), C&W (max iterations = 1000), and PGD ($\varepsilon = 0.1, \alpha = 0.01,$ and $I = 20$) attacks on cybersickness detection models. The notation X / Y represents the percentage score for the DeepTCN / Transformer model respectively.

| Attack Type | Acc. (%) | Precision (%) | Recall (%) | F1-S (%) |
|---|---|---|---|---|
| None | 91 / 95 | 92 / 96 | 94 / 98 | 94 / 97 |
| MI-FGSM | 51 / 43 | 45 / 40 | 47 / 42 | 46 / 41 |
| C&W | 22 / 16 | 24 / 17 | 22 / 16 | 23 / 17 |
| PGD | 34 / 26 | 30 / 24 | 31 / 27 | 28 / 24 |

Table 4: Transferability of adversarial attacks (e.g., MI-FGSM, PGD, and C&W) in the black-box setting. The notation X / Y / Z indicates model accuracy following MI-FGSM/PGD/C&W attacks, respectively.

| Generating Models | DeepTCN | Transformer |
|---|---|---|
| DeepTCN | - | 70 / 58 / 40 |
| Transformer | 74 / 63 / 46 | - |

## 6.1 Adversarial Attack Evaluation on MazeSick Dataset

This section presents the results of the impact of white-box and black-box cybersickness attacks on cybersickness detection models using the MazeSick dataset.

*(1) Impact of White-Box Adversarial Attacks in Cybersickness Detection Models:* To analyze the impact of adversarial cybersickness attacks, we generate adversarial examples using MI-FGSM ($\varepsilon = 0.1$), PGD ($\varepsilon = 0.1, \alpha = 0.01, I = 20$), and C&W (max iterations = 1000) attacks. We select these parameters based on established benchmarks for adversarial attacks targeting VR sensor data, as validated in prior work [18, 24]. The adversarial perturbations focus on both DeepTCN and Transformer models. Figures 5(a–c) illustrate adversarial examples crafted for the Transformer model. From Figures 5, we observe that PGD and C&W attacks produce adversarial examples closely resembling original VR sensor data. These stealthy attacks generally fall within the normal range of VR eye and head tracking data, which potentially induces VR cybersickness by triggering incorrect cybersickness mitigation strategies. Furthermore, to quantitatively assess the similarity between the original and adversarial examples, we calculate Pearson correlation coefficients (PCC) [3]. The analysis yields notably low PCC values, indicating minimal linear correlation between benign and adversarial data. For instance, the PCC values of C&W, PGD, and MI-FGSM adversarial attacks are 0.31, 0.29, and 0.23 for the Transformer model and 0.26, 0.23, and 0.21 for the DeepTCN models, respectively. These low correlations suggest that adversarial perturbations significantly alter the input data's statistical structure, thereby hindering the DL models' ability to generate consistent predictions.

Furthermore, Table 3 summarizes the performance of adversarial attacks on DeepTCN and Transformer models in terms of accuracy, precision, recall, and F1-score metrics. We observe that the MI-FGSM attack (with $\varepsilon = 0.1$) significantly decreases the cybersickness detection accuracy by approximately $1.78\times$ and $2.21\times$ for the DeepTCN and Transformer models, respectively. Moreover, the PGD ($\varepsilon = 0.1, \alpha = 0.01, I = 20$) and C&W (max iterations = 1000) decrease the accuracies of these cybersickness detection models to an even greater extent. Similar to the changes in accuracy, these attacks significantly affect the precision, recall, and F1 scores of these DL models, as shown in Table 3.

*(2) Impact of Black-Box Adversarial Attacks in Cybersickness Detection Models:* We conduct a systematic transferability analysis to evaluate the impact of black-box adversarial attacks on cybersickness detection models. Specifically, we apply adversarial examples generated for one cybersickness detection model to another. In black-box attack scenarios, the attacker lacks knowledge of the target model's internal parameters but can still significantly impact the target model's performance. The results of this transferability analysis are presented in Table 4. We observe that adversarial examples crafted for the Transformer model using MI-FGSM, PGD, and C&W attacks significantly reduce the DeepTCN model's accuracy from 91% without attack to 74%, 63%, and 46%, respectively. Similarly, adversarial examples crafted using DeepTCN reduce the Transformer model's accuracy from 95% without attack to 70%, 58%, and 40%, respectively. These findings clearly demonstrate that adversarial examples, particularly those created by more complex methods such as PGD and C&W, maintain their effectiveness across different cybersickness detection models. Thus, more complex adversarial attacks are generally more transferable across model types.

## 6.2 Adversarial Attack Evaluation on Testbed

This section provides a detailed evaluation of our deployed DL model within the proposed testbed, emphasizing the effectiveness of adversarial attacks and their impact on real-time cybersickness detection and mitigation. We utilized the trained Transformer model (trained on the MazeSick [20] dataset) for cybersickness detection tasks, as it demonstrated superior performance in cybersickness severity classification tasks, as shown in Table 3. The deployed Transformer model has a total size of approximately 53,320 KB, with a training duration of roughly 10 hours. During the VR simulation, the deployed Transformer model predicts cybersickness severity in 0.0021 seconds per frame, enabling near-instant predictions of cybersickness severity. During the experiment, we initiated adversarial attacks after 1 minute of VR simulation and sustained these attacks for 2 consecutive minutes to simulate realistic disruption scenarios. In this testbed, we restrict ourselves to white-box attack scenarios, where the adversary has complete knowledge of the cybersickness detection models, which enables the crafting of powerful adversarial examples. Figure 6 shows a participant immersed in the VR maze simulation under adversarial attack conditions, using our developed *Adversarial-VR* testbed. From Figure 6, it is observed that the deployed Transformer model detects the incorrect cybersickness severity class due to the attack (where an actual *"low"* severity class was incorrectly predicted as *"high"*), triggering the wrong mitigation technique and resulting in visual alterations that do not align with the user's actual cybersickness state. To further validate the impact of these adversarial attacks, we visualize predicted cybersickness severity levels for both without attack and adversarial conditions over a 120-second time window in the proposed testbed, as shown in Figure 7. We observe that in the attack condition, the prediction model frequently predicts the wrong cybersickness class. For example, at 105 seconds, the model incorrectly predicts *"high"* for an actual *"none"* severity class, triggering unnecessary strong visual tunneling mitigation (36°-80° FOV, AF = 0.52f). As a result, the mitigation layer restricted the user's FOV more than needed, despite the user not experiencing any cybersickness. This over-correction, caused by the adversarial attack, reduced visual comfort and interfered with the natural immersive experience. Note that overlapping points in the Figure 6 indicate cases where adversarial perturbations do not alter the model's prediction. However, we observe that in most instances, the adversarial attack successfully flips the model's prediction, except in a few cases where the model remains robust.

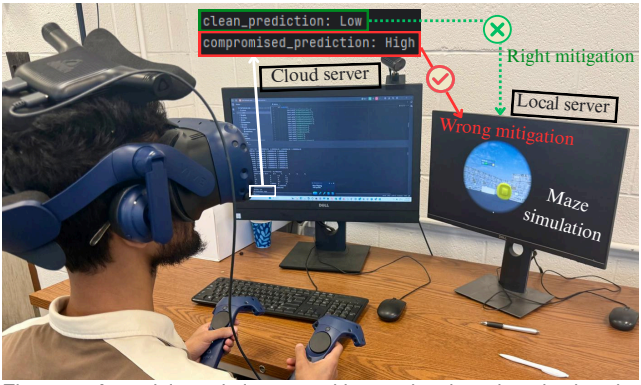

Figure 6: A participant is immersed in our developed testbed under adversarial attack conditions, where the DL-enabled model produces incorrect predictions due to adversarial perturbations, resulting in the activation of inappropriate mitigation strategies.

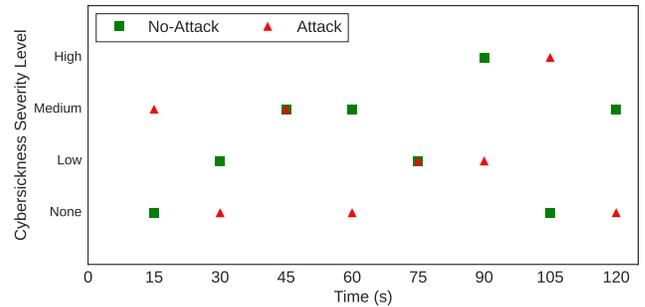

Figure 7: Comparison of predicted cybersickness severity levels under without attack and adversarial attack conditions over a 120-second VR simulation. The green squares represent accurate cybersickness severity predictions using benign (normal) data, while the red triangles illustrate adversarially manipulated predicted model outputs.

## 7 LIMITATIONS AND FUTURE WORK

Our proposed framework has a few limitations. First, the current implementation utilizes a controlled VR maze environment and only the MazeSick dataset, which lacks diverse VR environments (e.g., VR roller coasters) and demographic backgrounds (e.g., gender-imbalanced). While this setup is effective for demonstrating proof-of-concept vulnerabilities, it limits generalizability to broader VR experiences and populations. For future work, we plan to expand the testbed to include diverse environments (e.g., VR roller coasters, racing simulations, and interactive multiplayer scenarios) and incorporate datasets with equal gender representation and broader age groups to evaluate the robustness of AI-based cybersickness detection systems. Second, while we demonstrated vulnerability to three SOTA adversarial attacks (e.g., MI-FGSM, C&W, and PGD), future work will examine additional attack vectors, including advanced or hybrid strategies combining digital perturbations with physical-world sensor manipulations. Assessing these more complex threat models will provide deeper insights into realistic adversarial risks in VR environments. Lastly, due to time, resource, and budget constraints, we did not conduct a user study. This represents a limitation of our current work. Thus, in the future, we plan to conduct a comprehensive user study to validate the human-centered impacts of cybersickness and adversarial perturbations. Furthermore, we plan to explore defense mechanisms to detect and mitigate adversarial attacks, to strengthen the overall AI-based automatic cybersickness detection and mitigation framework.

## 8 CONCLUSION

This work presents a novel and open-source testbed specifically designed to evaluate adversarial threats to DL-based VR cybersickness detection and mitigation methods. To validate its effectiveness, we deployed the testbed on the HTC Vive Pro headsets, integrating DL-based models for cybersickness detection alongside a real-time dynamic FOV technique for mitigation. Specifically, we demonstrated that SOTA DL architectures, such as Transformer and DeepTCN models, are susceptible to advanced adversarial attacks (e.g., MI-FGSM, C&W, and PGD), which significantly degrade cybersickness detection accuracy and trigger unexpected mitigation strategies. For instance, our experimental results demonstrated that the C&W attack causes a $5.94\times$ decrease in accuracy for the Transformer-based cybersickness detection model, compared to the accuracy without the adversarial C&W attack. These vulnerabilities pose tangible risks to practical VR deployments, potentially undermining user comfort, immersion, and trust in the technology. By proactively identifying and addressing such threats and extending our testbed to support more diverse scenarios, sophisticated attack vectors, and robust defense strategies, the VR community can substantially strengthen the resilience and reliability of cybersickness detection systems, ultimately ensuring safer and more secure immersive experiences.

## 9 ACKNOWLEDGEMENT

This material is based upon work supported by the National Science Foundation (NSF) under Award Number CNS-2114035 and partially supported by the Army Research Offce (ARO) under award number W911NF-23-1-0401. Any opinions, findings, conclusions, or recommendations expressed in this publication are those of the authors and do not necessarily reflect the views of the NSF or ARO.